\documentclass{jps-cp}
\usepackage{txfonts} % Please comment out this line unless the txfonts package is availabe in your LaTeX system.

\title{Topological Characterization of Kitaev Spin Nanoribbons with Ordered Flux Configurations}

\author{Ryuto Tadokoro and Shoji Yamamoto}

\inst{Department of Physics, Hokkaido University, Sapporo 060-0810, Japan}

\email{yamamoto@phys.sci.hokudai.ac.jp}

\recdate{July 28, 2022}

\abst{
We demonstrate topological characterization of $S=\frac{1}{2}$ Kitaev quantum spin liquids on
a series of one-dimensional honeycomb nanoribbon lattices with zigzag and armchair terminated edges.
We draw their Majorana spinon phase diagrams with varying nearest-neighbor exchange couplings
not only at the sector of the ground flux configuration but also at some sectors of excited flux
configurations.
In the ground states of the zigzag and armchair nanoribbons, there occur a single and multiple phase
transitions, respectively, the former and latter of which are insensitive and subject to
the background flux configuration, respectively.
Topological phases each have a winding number as their invariant.
On each phase boundary, the Majorana spinon dispersion relation reflects both of
the change in the winding number and the background flux configuration.}

\kword{Kitaev quantum spin liquid, honeycomb nanoribbon,
       Majorana spinon, $\mathbb{Z}_{2}$ gauge flux,
       topological phase transition}

\begin{document}
\maketitle

\section{Introduction}

   The spin-$\frac{1}{2}$ Kitaev honeycomb model \cite{K2} is exactly solvable to have a spin-liquid
ground state \cite{B247201,S016502,K451,M012002}, whose elementary excitations are fractional,
decomposing into itinerant Majorana spinons and localized $\mathbb{Z}_{2}$ gauge fluxes.
Such spin-liquid states do not have any conventional long-range order and topological invariants
are proposed to identify them.
The Kitaev Hamiltonian is quadratic in Majorana fermions and block-diagonal with respect to
the emergent $\mathbb{Z}_{2}$ gauge flux configurations.
Topological properties of the gauge-ground (flux-free) sector have been theoretically studied well
\cite{K2,F087204,K207203}.
For instance, under the open boundary condition, there appear gapless Majorana excitation modes
\cite{T235434,M184418} on the boundaries according to the value of the topological invariant.
These ``Majorana edge modes" were indeed detected in the insulating two-dimensional quantum magnet
$\alpha$-RuCl$_3$ with bond-dependent Ising-type interactions
by observing its thermal Hall conductance \cite{K227}.

   On the other hand, gauge-excited sectors have been much less studied so far.
Recently, the Majorana correlations against several ordered flux configurations have been calculated
and turned out to be rather different from those in the flux-free sector \cite{K214421}.
There is a possibility of manipulating Majorana fermions by utilizing the gauge degrees of freedom
emergent in spin liquid states.
The Kitaev-spin-liquid candidate materials Na$_2$IrO$_3$ \cite{J045111},
$\alpha$-Li$_2$IrO$_3$ \cite{J025304}, and $\alpha$-RuCl$_3$ \cite{W3578} have been fabricated
into a thin film in recent years and further crystal shape tuning is under progress.
With all these in mind, we systematically study the one-dimensional Kitaev model on
a series of zigzag and armchair honeycomb nanoribbon lattices \cite{S012046}
with particular interest in whether and how their topological properties are affected by
the background gauge flux configuration.
There is a similarity between the Majorana excitation spectra of the Kitaev spin nanoribbons
and electron band structures of graphene nanoribbons \cite{E045432} indeed, while the gauge degrees
of freedom accompanying the former may cause further interest in \textit{tuning} the Majorana
excitation mechanism.

\section{Kitaev Models in Nanoribbon Geometry}

\begin{figure}[t]
\centering
\includegraphics[width=155mm]{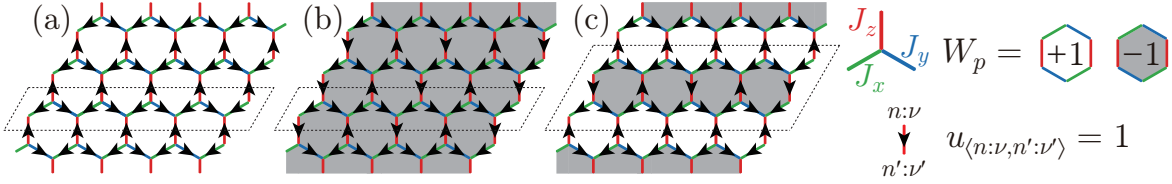}
\caption{Ordered flux configurations $\{W_{p}\}$:
         (a) the ground one, with $W_{p}=+1$ for all plaquettes (flux-free sector),
         and excited ones,
         (b) with $W_{p}=-1$ for all plaquettes (flux-full sector)
         and
         (c) alternating $W_{p}=\pm1$ (flux-half-occupied sector).
         The bond configurations $\{u_{\langle n:\nu,n':\nu'\rangle_{\lambda}}\}$
         yielding each flux configuration are also indicated by black arrows.
         We choose a unit cell as the dashed line.
         The case of the $(4,0)$-zigzag nanoribbon is shown as an example.}
\label{F:KitaevNanoribbonFlux}
\end{figure}

   The Kitaev Hamiltonian on nanoribbon lattices reads
\begin{align}
   \mathcal{H}
 =-\sum_{\lambda=x,y,z}\sum_{\langle n:\nu,n':\nu'\rangle_{\lambda}}
   J_{\lambda}\sigma_{n:\nu}^{\lambda}\sigma_{n':\nu'}^{\lambda},
   \label{E:Hspin}
\end{align}
where $(\sigma_{n:\nu}^{x},\sigma_{n:\nu}^{y},\sigma_{n:\nu}^{z})\ (n=1,\cdots,N;\ \nu=1,\cdots,R)$
are the Pauli matrices at the $\nu$th site in the $n$th unit,
while $\langle n:\nu,n':\nu'\rangle_{\lambda}$ runs over nearest-neighbor bonds
with $\lambda$ taking $x$, $y$, and $z$ (Fig. \ref{F:KitaevNanoribbonFlux}).
The coupling constants $J_{\lambda}$ are all set to $J_{\lambda}>0$ in the following.
The shape of a nanoribbon is specified by a set of two integers $(p,q)$ \cite{E045432},
hereafter referred to as the $(p,q)$-nanoribbon.
The case of $q=0$ and $q=1$ correspond to nanoribbons with zigzag and armchair terminated edges,
respectively, and $p$ merely adjusts the width of a nanoribbon.
We introduce four Majorana fermions at each site as
$\sigma_{n:\nu}^{\lambda}=i\eta_{n:\nu}^{\lambda}c_{n:\nu}$ with the anticommutation relations
$
 \{\eta_{n:\nu}^{\lambda},\eta_{n':\nu'}^{\lambda'}\}
=2\delta_{nn'}\delta_{\nu\nu'}\delta_{\lambda\lambda'}
$
,
$\{c_{n:\nu},c_{n':\nu'}\}=2\delta_{nn'}\delta_{\nu\nu'}$, and
$\{\eta_{n:\nu}^{\lambda},c_{n':\nu'}\}=0$ to obtain
\begin{align}
   \mathcal{H}
  =i\sum_{\lambda=x,y,z}\sum_{\langle n:\nu,n':\nu'\rangle_{\lambda}}
   J_{\lambda}\hat{u}_{\langle n:\nu,n':\nu'\rangle_{\lambda}}c_{n:\nu}c_{n':\nu'},
   \label{E:HMajorana}
\end{align}
where the nearest-neighbor bond operators
$
 \hat{u}_{\langle n:\nu,n':\nu'\rangle_{\lambda}}
 \equiv
 i\eta_{n:\nu}^{\lambda}\eta_{n':\nu'}^{\lambda}
=-\hat{u}_{\langle n':\nu',n:\nu\rangle_{\lambda}}
$
commute with each other as well as the Hamiltonian (\ref{E:HMajorana}) and therefore behave
as $\mathbb{Z}_{2}$ classical variables, $u_{\langle n:\nu,n':\nu'\rangle_{\lambda}}=\pm1$.
Note that four Majorana fermions at each site yield ``unphysical'' states \cite{Y217202,P165414},
which are projected out by the operator
$
\mathcal{P}
=
\prod_{n=1}^{N}\prod_{\nu=1}^{R}
\frac{1}{2}(1+\eta_{n:\nu}^{x}\eta_{n:\nu}^{y}\eta_{n:\nu}^{z}c_{n:\nu})
$
\cite{Y217202,P165414,Z014403}.
When we multiply six spin operators within each hexagon in the anticlockwise manner to define
the flux operator
\begin{align}
   \hat{W}_{p}
   \equiv
   \prod_{\langle n:\nu,n':\nu'\rangle_{\lambda}\in\partial p}
   \sigma_{n:\nu}^{\lambda}\sigma_{n':\nu'}^{\lambda}
  =-
   % (-i)^{6}
   \prod_{\langle n:\nu,n':\nu'\rangle_{\lambda}\in\partial p}
   \hat{u}_{\langle n:\nu,n':\nu'\rangle_{\lambda}},
   \label{E:Wp}
\end{align}
$\hat{W}_{p}$ also commutes with the Hamiltonian, whether (\ref{E:Hspin}) or (\ref{E:HMajorana}),
to behave as a classical variable, $W_{p}=\pm1$.
The Hilbert space of the Kitaev spin model (\ref{E:Hspin}) is block-diagonal with respect to
the flux configurations $\{W_{p}\}$.
Given a set of the bond variables $\{u_{\langle n:\nu,n':\nu'\rangle_{\lambda}}\}$ yielding
a certain flux configuration $\{W_{p}\}$, the augmented Majorana Hamiltonian (\ref{E:HMajorana})
becomes a Majorana quadratic form in such flux configuration sector.
The eigenspectrum of (\ref{E:HMajorana}) depends on $\{u_{\langle n:\nu,n':\nu'\rangle_{\lambda}}\}$
only through $\{W_{p}\}$.

   We consider three types of ordered flux configurations, i.e.,
flux-free (gauge-ground) [Fig. \ref{F:KitaevNanoribbonFlux}(a)],
flux-full [Fig. \ref{F:KitaevNanoribbonFlux}(b)], and
flux-half-occupied [Fig. \ref{F:KitaevNanoribbonFlux}(c)] sectors.
For all these flux configurations, we have the Majorana quadratic Hamiltonian
\begin{align}
   \mathcal{H}
  =\sum_{k=k_{1}}^{k_{N}}\boldsymbol{c}_{k}^{\dagger}H(k)\boldsymbol{c}_{k};
   \qquad
   H(k)
  =\frac{i}{2}
   \begin{bmatrix}
     0              & h(k) \\
    -h^{\dagger}(k) & 0
   \end{bmatrix}
   \label{E:MatrixH}
\end{align}
in the momentum space,
where $h(k)$ is the complex matrix of dimension $\frac{R}{2}\times\frac{R}{2}$ with $R$
depending on the background flux configuration $\{W_{p}\}$ as well as $p$ and $q$.
The gauge-fixed quadratic Hamiltonian (\ref{E:MatrixH}) is diagonalized into
$
 \mathcal{H}
=\sum_{k=k_{1}}^{k_{N}}\sum_{\mu=1}^{R/2}\varepsilon_{k:\mu}
 \left(
 \alpha_{k:\mu}^{\dagger}\alpha_{k:\mu}-\frac{1}{2}
 \right),
 \label{E:HDiag}
$
where the eigenvalues $\varepsilon_{k:\mu}$ are the singular values of $h(k)$
\cite{Z014403,W115146} and therefore nonnegative.
We understand that the primitive translation be unity in each nanoribbon throughout the manuscript.

\section{Topological Phase Transitions}

   We discuss topological phase transitions of the Kitaev nanoribbons with ordered flux configurations.
This model belongs to the BDI symmetry class within the ten-fold way classification
\cite{S195125,R065010}, and the integer-valued topological invariant for gapped phases
is given by the winding number
\begin{align}
   \nu_{\mathrm{w}}
  =\frac{1}{2\pi i}\int_{-\pi}^{\pi}dk
   \frac{\partial}{\partial k}\ln\det h(k),
   \label{E:WN}
\end{align}
where $h(k)$ is the off-diagonal box of Eq. (\ref{E:MatrixH}).
In this expression, $\nu_{\mathrm{w}}$ corresponds to the number of times $\det h(k)$ winds around
the origin of the complex plane as $k$ is varied from $-\pi$ to $\pi$.
According to the bulk-edge correspondence, the topological phase exhibits
gapless Majorana edge modes, the number of which is equal to $|\nu_{\mathrm{w}}|$ \cite{V057001}.
Note that if the Majorana spinon excitation spectrum is gapless,
then $|\det h(k)|=\prod_{\mu=1}^{R/2}\varepsilon_{k:\mu}=0$,
and therefore the winding number can no longer be well-defined.
In order to overcome this difficulty and obtain a more tractable expression of the winding number,
we put $z=e^{ik}$ in Eq. (\ref{E:WN})
and perform an analytic continuation of $\det h(k)$ to the entire complex plane.
Abusing notation, we write $\det h(k)\to\det h(z)$, where $\det h(z)$ becomes a polynomial in $z$.
Then we can calculate the winding number using Cauchy's argument principle as
\begin{align}
   \nu_{\mathrm{w}}
  =N_{\mathrm{zero}}^{|z|<1}-N_{\mathrm{pole}}^{|z|<1},
   \label{E:ExtendedWN}
\end{align}
where $N_{\mathrm{zero}}^{|z|<1}\ (N_{\mathrm{pole}}^{|z|<1})$ is the number of zeros (poles) of
$\det h(z)$ in $|z|<1$ counted as many times its multiplicity (order).
$N_{\mathrm{zero}}^{|z|<1}-N_{\mathrm{pole}}^{|z|<1}$ can always be calculated regardless of
whether the Majorana spinon excitation spectrum is gapped or gapless.
Topological phase transitions, i.e., the winding number changes,
are caused by zeros moving into or out of the unit circle.
On the phase boundary, with the zero $z_{0}=e^{ik_{0}}$ on the unit circle,
the Majorana spinon excitation spectrum becomes gapless at $k=k_{0}$
and behaves as $\varepsilon_{k:\mu}\propto |k-k_{0}|^{m}\ (|k-k_{0}|\ll 1)$,
where $m$ denotes the multiplicity of the zero \cite{V057001}.

\begin{figure}[t]
\centering
\includegraphics[width=153mm]{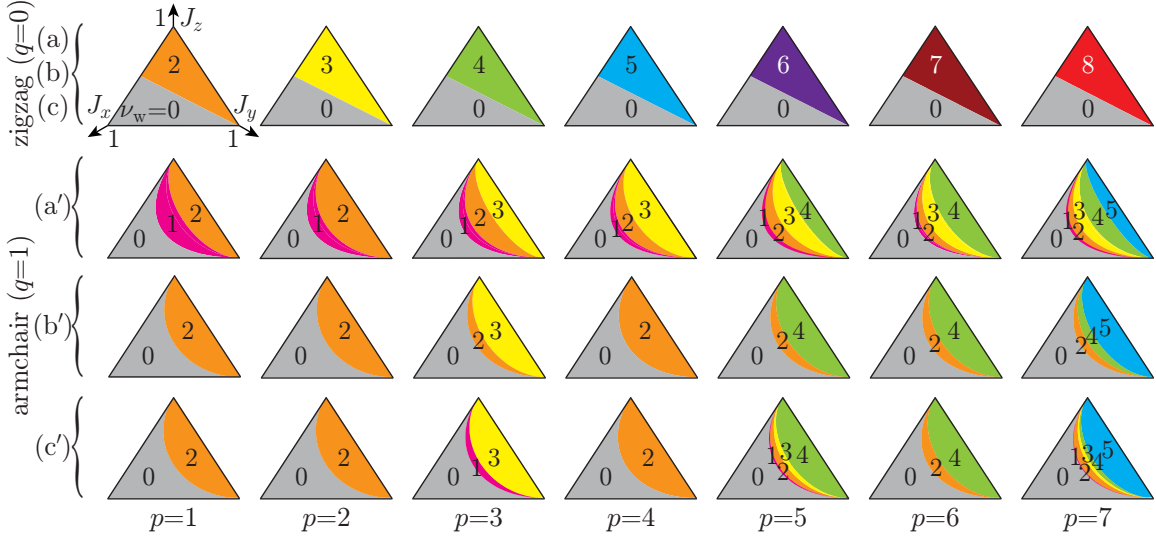}
\caption{Topological phase diagrams of the $(p,0)$-zigzag [(a) to (c)] and
         $(p,1)$-armchair [($\mathrm{a}^{\prime}$) to ($\mathrm{c}^{\prime}$)]
         nanoribbons with varying ribbon width $p$ on the plane $J_{x}+J_{y}+J_{z}=1$,
         each labeled with various winding numbers $\nu_{\mathrm{w}}$.
         The three types of ordered flux configurations depicted in
         Fig. \ref{F:KitaevNanoribbonFlux} are calculated in both types of nanoribbons.}
\label{F:TPDs}
\end{figure}

   Figure \ref{F:TPDs} shows topological phase diagrams obtained
by numerically evaluating Eq. (\ref{E:WN}).
In the case of the zigzag nanoribbons, exactly the same topological phase diagram---a single transition
between the $\nu_{\mathrm{w}}=0$ trivial phase and a $\nu_{\mathrm{w}}\ne0$ topological phase
at $J_{x}=J_{z}$---is obtained for all the flux configurations.
On the other hand, topological phase diagrams of the armchair nanoribbons exhibit a multiple transition
and vary with the background flux configuration.

\begin{figure}[t]
\centering
\includegraphics[width=153mm]{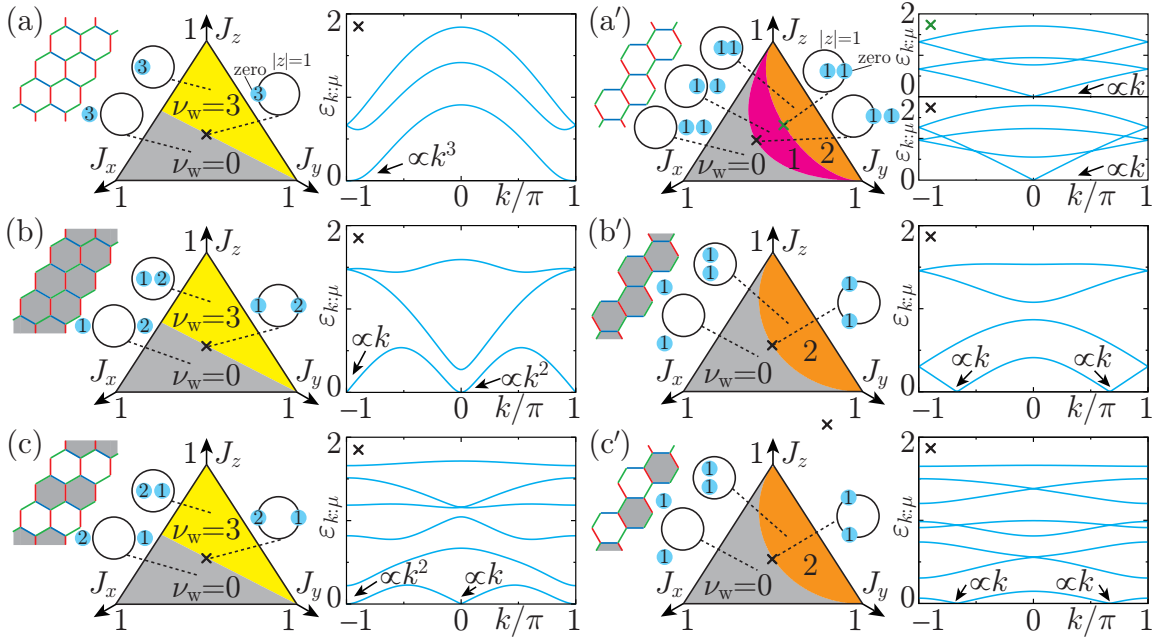}
\caption{Topological phase diagrams on the $J_{x}+J_{y}+J_{z}=1$ plane of
         the $(2,0)$-zigzag [(a) to (c)] and $(1,1)$-armchair nanoribbons
         [($\mathrm{a}^{\prime}$) to ($\mathrm{c}^{\prime}$)]
         in various flux configulations.
         Figures covered with blue circles denote the \textit{zeros} (with multiplicity in general)
         of $\det h(z)$ in the complex plane.
         They are exactly on the unit circle centered at the origin when they cross
         a phase boundary, where
         the Majorana spinon dispersion relations $\varepsilon_{k:\mu}\ (\mu=1,\cdots,\frac{R}{2})$
         exhibit corresponding \textit{nodes}.}
\label{F:TPD&Ek}
\end{figure}

   Let us investigate the ground-flux-configuration (flux-free) sector.
$h(z)$ for the zigzag nanoribbons becomes a lower triangular matrix of dimension $(p+1)\times(p+1)$,
whose nonzero elements consist of diagonal and $(\mu+1,\mu)$-off-diagonal ones being
$J_{x}+J_{z}z$ and $J_{y}$, respectively.
Its determinant reads
\begin{align}
   \det h_{\mathrm{GS}}^{\mathrm{Z}}(z)
  =\begin{vmatrix}
    J_{x}+J_{z}z & 0            & \cdots & \cdots & \cdots       & 0            \\
    J_{y}        & J_{x}+J_{z}z & \ddots &        &              & \vdots       \\
    0            & J_{y}        & \ddots & \ddots &              & \vdots       \\
    \vdots       & \ddots       & \ddots & \ddots & \ddots       & \vdots       \\
    \vdots       &              & \ddots & J_{y}  & J_{x}+J_{z}z & 0            \\
    0            & \cdots       & \cdots & 0      & J_{y}        & J_{x}+J_{z}z
   \end{vmatrix}
  =(J_{x}+J_{z}z)^{p+1}
   \label{E:dethGS}
\end{align}
and yields the zero $z=-J_{x}/J_{z}$ of multiplicity $p+1$ without any pole.
We thus obtain the same type of topological phase diagram which consists of $\nu_{\mathrm{w}}=0$
and $\nu_{\mathrm{w}}=p+1$ phases, corresponding to the case where the zero exists
inside and outside the unit circle, respectively [Figs. \ref{F:TPDs}(a) and \ref{F:TPD&Ek}(a)].
Since the excitation gap vanishes when the zero exists on the unit circle, we find
a gap-closing condition $J_{x}=J_{z}$ by solving $|z|=1$, which represents the phase boundary.
A single topological phase transition occurs when the zero passes over $z=e^{i\pi}=-1$,
as is shown in Fig. \ref{F:TPD&Ek}(a).
The Majorana spinon excitation spectrum on the phase boundary thus becomes gapless at $k=\pi$ and
behaves as $\varepsilon_{k:\mu}\propto|k-\pi|^{p+1}\ (|k-\pi|\ll 1)$.
In the armchair nanoribbons, on the other hand, $h(z)$ becomes a tridiagonal Toeplitz
matrix of dimension $(p+3)\times(p+3)$,
whose diagonal, $(\mu+1,\mu)$-off-diagonal, and $(\mu,\mu+1)$-off-diagonal elements are
$J_{x}$, $J_{y}$, and $J_{z}z$, respectively.
Its determinant reads
\begin{align}
   \det h_{\mathrm{GS}}^{\mathrm{A}}(z)
  =\begin{vmatrix}
    J_{x}  & J_{z}z & 0      & \cdots & \cdots & 0      \\
    J_{y}  & J_{x}  & J_{z}z & \ddots &        & \vdots \\
    0      & J_{y}  & \ddots & \ddots & \ddots & \vdots \\
    \vdots & \ddots & \ddots & \ddots & J_{z}z & 0      \\
    \vdots &        & \ddots & J_{y}  & J_{x}  & J_{z}z \\
    0      & \cdots & \cdots & 0      & J_{y}  & J_{x}
   \end{vmatrix}
  =\prod_{i=1}^{p+3}
   \left[J_{x}+2\sqrt{J_{y}J_{z}z}\cos{\left(\frac{i\pi}{p+4}\right)}\right]
   \label{E:dethGS_A}
\end{align}
and yields the simple zeros
$
 z
=\frac{(J_{x})^{2}}{4J_{y}J_{z}\cos^{2}{(\frac{i\pi}{p+4})}}
 \ (i=1,\cdots,\lfloor\frac{p+3}{2}\rfloor)
$
without any pole, where $\lfloor x\rfloor$ is the floor function giving the greatest integer
that is less than or equal to $x$.
The simple zeros move into or out of the unit circle one by one, resulting in a multiple transition
where the winding number varies one by one from $\nu_{\mathrm{w}}=0$ to
$\nu_{\mathrm{w}}=\lfloor\frac{p+3}{2}\rfloor$
[Figs. \ref{F:TPDs}($\mathrm{a}^{\prime}$) and \ref{F:TPD&Ek}($\mathrm{a}^{\prime}$)].
Topological phase transitions occur when the simple zeros pass over $z=e^{i\cdot0}=1$,
as is shown in Fig. \ref{F:TPD&Ek}($\mathrm{a}^{\prime}$).
Thus in the armchair nanoribbons, the Majorana spinon excitation spectrum on the phase boundary
becomes gapless at $k=0$ with a linear dispersion relation.
It is also the case with the zigzag nanoribbons
but at $k=\pi$ with a quadratic or higher dispersion relation.

   We further investigate the excited-flux-configuration sectors intending to reveal
the effect of the background flux configuration on the topological phase diagrams.
First we discuss the zigzag nanoribbons.
The determinants of $h(z)$ in the flux-full [Fig. \ref{F:KitaevNanoribbonFlux}(b)]
and flux-half-occupied [Fig. \ref{F:KitaevNanoribbonFlux}(c)] sectors read
\begin{align}
   \det h_{\mathrm{ES1}}^{\mathrm{Z}}(z)
   &
  =(J_{x}-J_{z}z)^{\left\lfloor\frac{p+2}{2}\right\rfloor}
   (J_{x}+J_{z}z)^{\left\lfloor\frac{p+1}{2}\right\rfloor},
   \label{E:dethES1}
   \\
   \det h_{\mathrm{ES2}}^{\mathrm{Z}}(z)
   &
  =(J_{x})^{p+1}
   [J_{x}+(J_{z})^{2}z/J_{x}]^{\left\lfloor\frac{p+2}{2}\right\rfloor}
   [J_{x}-(J_{z})^{2}z/J_{x}]^{\left\lfloor\frac{p+1}{2}\right\rfloor},
   \label{E:dethES2}
\end{align}
respectively.
Unlike the case of the ground flux configuration, two zeros are obtained
in each excited flux configuration, i.e.,
$z=\pm J_{x}/J_{z}$ from Eq. (\ref{E:dethES1}) and
$z=\pm(J_{x})^{2}/(J_{z})^{2}$ from Eq. (\ref{E:dethES2}).
We find not only the same gap-closing condition $J_{x}=J_{z}$ as the ground state
but also the same values of the winding number, $\nu_{\mathrm{w}}=0$ and
$\nu_{\mathrm{w}}=\left\lfloor\frac{p+1}{2}\right\rfloor+\left\lfloor\frac{p+2}{2}\right\rfloor=p+1$.
Note that a topological phase transition occurs when the two zeros of multiplicity
$\left\lfloor\frac{p+1}{2}\right\rfloor$ and $\left\lfloor\frac{p+2}{2}\right\rfloor$
move simultaneously into or out of the unit circle at $z=\pm1$
[Figs. \ref{F:TPD&Ek}(b) and \ref{F:TPD&Ek}(c)].
Hence the Majorana spinon excitation spectrum on the phase boundary becomes gapless at $k=0$
as well as at $k=\pi$.
They behave as
$\varepsilon_{k:\mu}\propto |k-\pi|^{\left\lfloor\frac{p+1}{2}\right\rfloor}\ (|k-\pi|\ll 1)$ and
$\varepsilon_{k:\mu}\propto |k|^{\left\lfloor\frac{p+2}{2}\right\rfloor}\ (|k|\ll 1)$
in the flux-full sector [Fig. \ref{F:TPD&Ek}(b)],
while the behavior at $k=0$ and $k=\pi$ are reversed in the flux-half-occupied sector
[Fig. \ref{F:TPD&Ek}(c)].
The topological phase diagrams in the gauge-excited sectors are exactly the same as those
in the gauge-ground sector, but the Majorana spinon dispersion relations
vary with their background gauge configurations.
Next we discuss the armchair nanoribbons, even though they are less analyzable.
Numerical findings generally show that the number of phase transitions with excited flux
configurations is generally smaller than that in the ground flux configuration
[Figs. \ref{F:TPDs}($\mathrm{a}^{\prime}$)--\ref{F:TPDs}($\mathrm{c}^{\prime}$)].
The zeros cross the unit circle one by one at the same point in the ground flux configuration,
whereas they can cross the unit circle simultaneously at different points in general against
an excited flux configuration.
% [Fig. \ref{F:TPD&Ek}($\mathrm{a}^{\prime}$)--\ref{F:TPD&Ek}($\mathrm{c}^{\prime}$)].
However, unlike the zigzag nanoribbons, all the zeros cross the unit circle \textit{alone}.
No node is degenerate, and therefore, the Majorana spinon excitation spectrum on the phase boundary
is always characterized by one or more \textit{linear} dispersion relations.

\section{Concluding Remarks}

   How many and what kind of topological phases occur with varying anisotropic exchange couplings
are insensitive and subject to the background gauge flux configuration
in the zigzag and armchair nanoribbons, respectively.
Every time we cross a phase boundary, mode softening occurs in the Majorana spinon dispersion
relation at one or more particular values of momentum.
While the gapless dispersion relations depend not only on the jump in the topological invariant
but also on the background gauge flux configuration,
the zigzag and armchair nanoribbons seem to show a remarkable difference in this context.
With various background flux configurations,
mode softening in the former may be accompanied by any integral order of dispersion relation,
whereas that in the latter is always accompanied by a linear dispersion relation.

   Similar to graphene nanoribbons \cite{T205311} and carbon nanotubes \cite{I195442},
the present model belongs to the one-dimensional BDI symmetry class \cite{S195125,R065010}
but its gauge degrees of freedom distinguishes itself from the others.
A similar analysis in lower and higher dimensions is encouraged, i.e.,
topological characterization of
Kitaev spin balls \cite{K214411} and a $\mathbf{C}_{4\mathrm{v}}$ Kitaev spin plane
\cite{Y063701}, for instance.

\vspace{\baselineskip}
   This work is supported by JST SPRING Grant No. JPMJSP2119 and
JSPS KAKENHI Grant No. 22K03502.


\begin{thebibliography}{99}
\bibitem{K2}
   A. Kitaev,
      Ann. Phys. (N.Y.) \textbf{321}, 2 (2006).

\bibitem{B247201}
   G. Baskaran, S. Mandal, and R. Shankar,
      Phys. Rev. Lett. \textbf{98}, 247201 (2007).

\bibitem{S016502}
   L. Savary and L. Balents,
      Rep. Prog. Phys. \textbf{80}, 016502 (2017).

\bibitem{K451}
   J. Knolle and R. Moessner,
      Annu. Rev. Condens. Matter Phys. \textbf{10}, 451 (2019).

\bibitem{M012002}
   Y. Motome and J. Nasu,
      J. Phys. Soc. Jpn. \textbf{89}, 012002 (2020).

\bibitem{F087204}
   X.-Y. Feng, G.-M. Zhang, and T. Xiang,
      Phys. Rev. Lett. \textbf{98}, 087204 (2007).

\bibitem{K207203}
   J. Knolle, D. L. Kovrizhin, J. T. Chalker, and R. Moessner,
      Phys. Rev. Lett. \textbf{112}, 207203 (2014).

\bibitem{T235434}
   M. Thakurathi, K. Sengupta, and D. Sen,
      Phys. Rev. B \textbf{89}, 235434 (2014).

\bibitem{M184418}
   T. Mizoguchi and T. Koma,
      Phys. Rev. B \textbf{99}, 184418 (2019).

\bibitem{K227}
   Y. Kasahara, T. Ohnishi, Y. Mizukami, O. Tanaka, S. Ma, K. Sugii, N. Kurita,
   H. Tanaka, J. Nasu, Y. Motome, T. Shibauchi, and Y. Matsuda,
      Nature (London) \textbf{559}, 227 (2018).

\bibitem{K214421}
   A. Koga, Y. Murakami, and J. Nasu,
      Phys. Rev. B \textbf{103}, 214421 (2021).

\bibitem{J045111}
   M. Jenderka, J. Barzola-Quiquia, Z. Zhang, H. Frenzel, M. Grundmann, and M. Lorenz,
      Phys. Rev. B \textbf{88}, 045111 (2013).

\bibitem{J025304}
   M. Jenderka, R. Schmidt-Grund, M. Grundmann, and M. Lorenz,
      J. Appl. Phys. \textbf{117}, 025304 (2015).

\bibitem{W3578}
   D. Weber, L. M. Schoop, V. Duppel, J. M. Lippmann, J. Nuss, and B. V. Lotsch,
      Nano Lett. \textbf{16}, 3578 (2016).

\bibitem{S012046}
   K. Suzuki and S. Yamamoto,
      J. Phys.: Conf. Ser. \textbf{1220}, 012046 (2019).

\bibitem{E045432}
   M. Ezawa,
      Phys. Rev. B \textbf{73}, 045432 (2006).

\bibitem{Y217202}
   H. Yao, S.-C. Zhang, and S. A. Kivelson,
      Phys. Rev. Lett. \textbf{102}, 217202 (2009).

\bibitem{P165414}
   F. L. Pedrocchi, S. Chesi, and D. Loss,
      Phys. Rev. B \textbf{84}, 165414 (2011).

\bibitem{Z014403}
   F. Zschocke and M. Vojta,
      Phys. Rev. B \textbf{92}, 014403 (2015).

\bibitem{W115146}
   A, J. Willans, J. T. Chalker, and R. Moessner,
      Phys. Rev. B \textbf{84}, 115146 (2011).

\bibitem{S195125}
   A. P. Schnyder, S. Ryu, A. Furusaki, and A. W. W. Ludwig,
      Phys. Rev. B \textbf{78}, 195125 (2008).

\bibitem{R065010}
   S. Ryu, A. P. Schnyder, A. Furusaki, and A. W. W. Ludwig,
      New J. Phys. \textbf{12}, 065010 (2010).

\bibitem{V057001}
   R. Verresen, N. G. Jones, and F. Pollmann,
      Phys. Rev. Lett. \textbf{120}, 057001 (2018).

\bibitem{T205311}
   G. Tamaki, T. Kawakami, and M. Koshino,
      Phys. Rev. B \textbf{101}, 205311 (2020).

\bibitem{I195442}
   W. Izumida, R. Okuyama, A. Yamakage, and R. Saito,
      Phys. Rev. B \textbf{93}, 195442 (2016).

\bibitem{K214411} % Kitaev spin ball
   T. Kimura and S. Yamamoto,
      Phys. Rev. B \textbf{101}, 214411 (2020).

\bibitem{Y063701} % 2D Kitaev series, depolarization
   S. Yamamoto and T. Kimura,
      J. Phys. Soc. Jpn. \textbf{89}, 063701 (2020).

\end{thebibliography}
\end{document}